# A fast and long-lived outflow from the supermassive black hole in NGC 5548

**Authors:** J.S. Kaastra[1,2,3]*, G.A. Kriss[4,5], M. Cappi[6], M. Mehdipour[1,7], P.-O. Petrucci[8,9], K.C. Steenbrugge[10,11], N. Arav[12], E. Behar[13], S. Bianchi[14], R. Boissay[15], G. Branduardi-Raymont[7], C. Chamberlain[12], E. Costantini[1], J.C. Ely[4], J. Ebrero[1,16], L. Di Gesu[1], F.A. Harrison[17], S. Kaspi[13], J. Malzac[18,19], B. De Marco[20], G. Matt[14], K. Nandra[20], S. Paltani[15], R. Person[21], B.M. Peterson[22,23], C. Pinto[24], G. Ponti[20], F. Pozo Nuñez[25], A. De Rosa[26], H. Seta[27], F. Ursini[8,9], C.P. de Vries[1], D.J. Walton[17], M. Whewell[7].

**Affiliations:**

**[1]**SRON Netherlands Institute for Space Research, Sorbonnelaan 2, 3584 CA Utrecht, the Netherlands.

**[2]**Department of Physics and Astronomy, Universiteit Utrecht, P.O. Box 80000, 3508 TA Utrecht, the Netherlands.

[3]Leiden Observatory, Leiden University, PO Box 9513, 2300 RA Leiden, the Netherlands.

[4]Space Telescope Science Institute, 3700 San Martin Drive, Baltimore, MD 21218, USA.

**[5]**Department of Physics and Astronomy, The Johns Hopkins University, Baltimore, MD 21218, USA.

**[6]**INAF-IASF Bologna, Via Gobetti 101, I-40129 Bologna, Italy.

**[7]**Mullard Space Science Laboratory, University College London, Holmbury St. Mary, Dorking, Surrey, RH5 6NT, UK.

**[8]**Univ. Grenoble Alpes, IPAG, F-38000 Grenoble, France.

[9]CNRS, IPAG, F-38000 Grenoble, France.

**[10]**Instituto de Astronomía, Universidad Católica del Norte, Avenida Angamos 0610, Casilla 1280, Antofagasta, Chile.

**[11]**Department of Physics, University of Oxford, Keble Road, Oxford, OX1 3RH, UK.

**[12]**Department of Physics, Virginia Tech, Blacksburg, VA 24061, USA.

**[13]**Department of Physics, Technion-Israel Institute of Technology, 32000 Haifa, Israel.

**[14]**Dipartimento di Matematica e Fisica, Università degli Studi Roma Tre, via della Vasca Navale 84, 00146 Roma, Italy.

**[15]**Department of Astronomy, University of Geneva, 16 Ch. d'Ecogia, 1290 Versoix, Switzerland.

[16]European Space Astronomy Centre (ESAC), P.O. Box 78, E-28691 Villanueva de la Cañada, Madrid, Spain.

[17]Cahill Center for Astronomy and Astrophysics, California Institute of Technology, Pasadena, CA 91125, USA.

[18]Université de Toulouse, UPS-OMP, IRAP, Toulouse, France.

[19]CNRS, IRAP, 9 Av. colonel Roche, BP 44346, 31028 Toulouse Cedex 4, France.

[20]Max-Planck-Institut für extraterrestrische Physik, Giessenbachstrasse, D-85748 Garching, Germany.

[21]22 Impasse du Bois Joli, 74410 St. Jorioz, France.

[22]Department of Astronomy, The Ohio State University, 140 W 18th Avenue, Columbus, OH 43210, USA.

[23]Center for Cosmology & AstroParticle Physics, The Ohio State University, 191 West Woodruff Avenue, Columbus, OH 43210, USA.

[24]Institute of Astronomy, University of Cambridge, Madingley Rd, Cambridge, CB3 0HA, UK.

[25]Astronomisches Institut, Ruhr-Universität Bochum, Universitätsstraße 150, 44801, Bochum, Germany.

[26]INAF/IAPS - Via Fosso del Cavaliere 100, I-00133 Roma, Italy.

[27]Research Center for Measurement in Advanced Science, Faculty of Science, Rikkyo University 3-34-1 Nishi-Ikebukuro, Toshima-ku, Tokyo, Japan.

*Correspondence to: J.Kaastra@sron.nl.

**Abstract**: Supermassive black holes in the nuclei of active galaxies expel large amounts of matter through powerful winds of ionized gas. The archetypal active galaxy NGC 5548 has been studied for decades, and high-resolution X-ray and UV observations have previously shown a persistent ionized outflow. An observing campaign in 2013 with six space observatories shows the nucleus to be obscured by a long-lasting, clumpy stream of ionized gas never seen before. It blocks 90% of the soft X-ray emission and causes simultaneous deep, broad UV absorption troughs. The outflow velocities of this gas are up to five times faster than those in the persistent outflow, and at a distance of only a few light days from the nucleus, it may likely originate from the accretion disk.

Outflows of photo-ionized gas are ubiquitous in active galactic nuclei (AGN) such as Seyfert galaxies (*1, 2*). Their impact on the environment can be estimated from the mass loss rate per solid angle, which is proportional to the hydrogen column density $N_H$, outflow velocity $v$ and distance $r$ to the ionizing source. While the first two quantities are directly obtained from high-resolution spectral observations, in most cases the distance can only be inferred from the ionization parameter $\xi = L/nr^2$, with $n$ being the hydrogen density and $L$ the ionizing luminosity between 13.6 eV and 13.6 keV. The density can be obtained from density-sensitive UV lines or by measuring the recombination time scale $t_{rec} \sim 1/n$ by monitoring the response of the outflow to continuum variations. The latter approach has recently been successfully applied to Mrk 509, providing strong constraints on the distances of the five ionization components in that source (*3, 4*).

Motivated by the successful results of the Mrk 509 campaign, we have conducted comprehensive monitoring of NGC 5548, an archetypal Seyfert 1 galaxy ($z$=0.017175). This campaign from May 2013 to February 2014 involved observations with the XMM-Newton, Hubble Space telescope (HST), Swift, the Nuclear Spectroscopic Telescope Array (NuSTAR), the INTErnational Gamma-Ray Astrophysics Laboratory (INTEGRAL), and Chandra satellites (*5*).

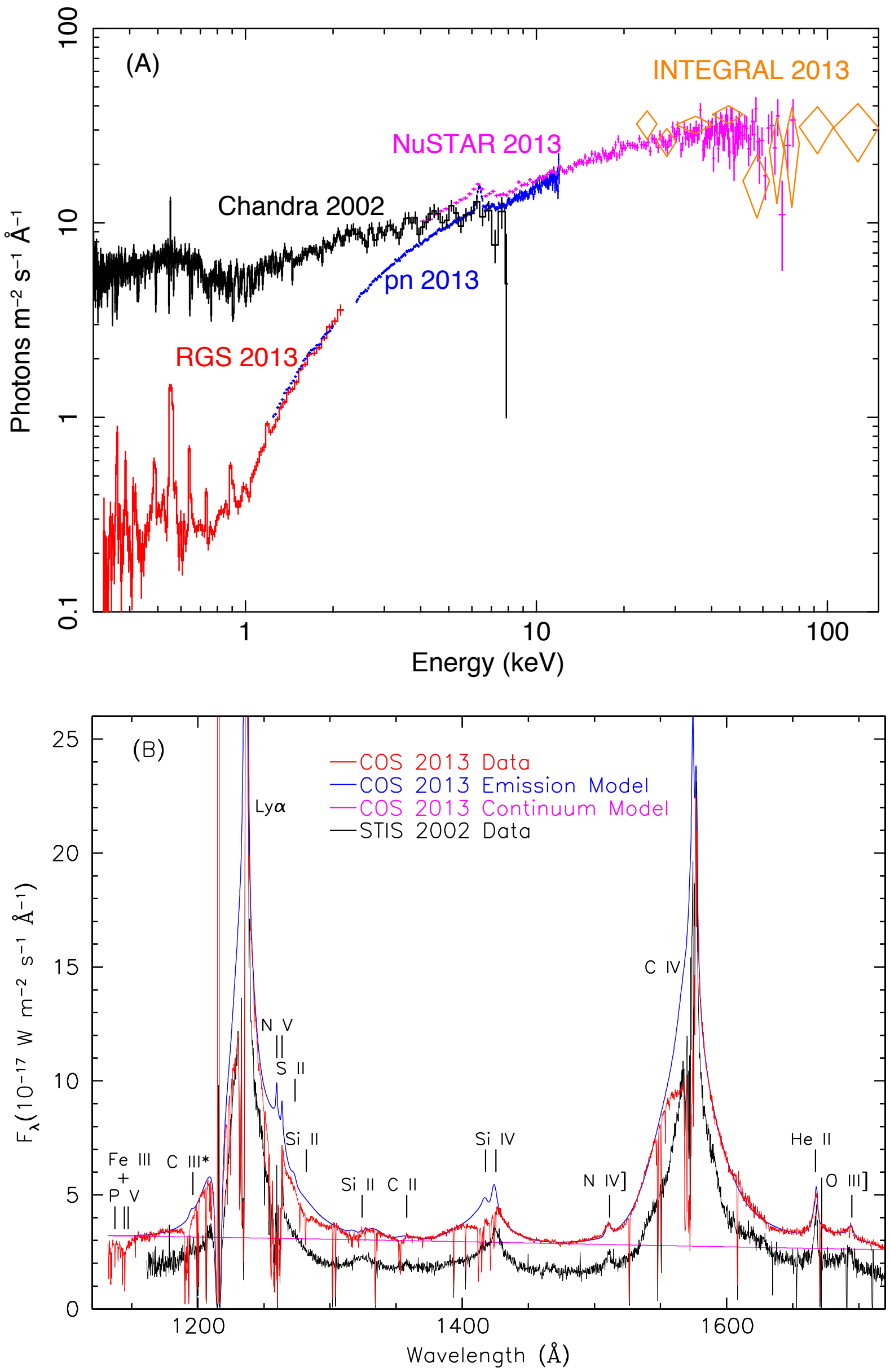

**Fig. 1.** X-ray and UV spectra of NGC 5548. All data have been rebinned for clarity. Error bars are ±1 SD. (A) The heavily obscured X-ray spectrum during summer 2013. The unobscured Chandra Low Energy Transmission Grating Spectrometer (LETGS) spectrum taken in 2002 is shown for comparison. The 2013 spectrum is obtained from 12 XMM-Newton observations (EPIC-pn detector (pn) and Reflection Grating Spectrometers (RGS)), two NuSTAR and four INTEGRAL observations; these latter two datasets were taken when the hard X-ray flux was 10% higher than the average >10 keV flux of the XMM-Newton data. (B) Averaged 2013 Cosmic Origins Spectrograph (COS) spectrum compared to 2002 Space Telescope Imaging Spectrograph (STIS) spectrum, showing the broad UV absorption lines in 2013.

To our surprise, the soft X-ray flux in the first XMM-Newton observation (22 June 2013) was 25 times weaker than the typical median as measured with Chandra LETGS in 2002, and this strong suppression was consistent throughout all 14 XMM-Newton spectra (Fig. 1). The NuSTAR and INTEGRAL spectra at energies above 10 keV show the characteristic power-law shape with a photon index of 1.6–1.7 and a weak constant reflection component, consistent with the median flux level of the 2002 Chandra/LETGS observation. However, the spectrum is cut off below 10 keV, reaching an effective photon index of −0.5 in the 1–2 keV band, before it flattens again below 1 keV. At energies <1 keV, the spectrum shows clear narrow emission lines and radiative recombination continua from photoionized gas at a temperature of ~6 eV. The intensity of the strongest emission line, the [O VII] forbidden line at 0.56 keV, is only slightly below its 2002 intensity. These narrow features are superimposed on a weak continuum.

Comptonized spectra cannot explain the observed hard photon index. The central continuum is more likely absorbed by obscuring material that partially covers (~90%) the source. The absorbing gas must be inside the X-ray narrow line region; we concluded that this region is not obscured, because the intensity of the narrow emission features is not suppressed compared to archival data.

We obtained HST/COS spectra concurrently with six of our 14 XMM-Newton spectra (Fig. 1). These UV spectra show a large number of narrow absorption lines at the velocities of the classical warm absorber (WA) components that were already present 20 years ago (*1*), but which now originate from less-ionized gas. This shows that the WA is now irradiated by a much lower ionizing flux than before.

After modeling the emission lines and continuum and accounting for the narrow absorption lines (*5*), broad, blue-shifted absorption troughs remain visible in the COS spectra (Fig. 2). Asymmetric troughs reach their deepest point at outflow velocities of −1000 km/s and extend to blueshifts as high as −5000 km/s. All permitted UV transitions in the COS spectra show this blueshifted absorption: C II, Si II, Si II*, S II, Fe III, C III*, Si III, Si IV, C IV, N V, and Lyα. As expected, no absorption is seen associated with forbidden or semi-forbidden transitions, or highly excited states such as He II λ1640. As we show below, all visible broad absorption transitions appear in ions formed at ionization parameters similar to those needed to explain the X-ray obscuration. In addition, the strengths of the broad absorption troughs in the individual observations are correlated with the strength of the X-ray obscuration (Fig. 2C). This indicates that the broad UV absorption troughs and the X-ray obscuration arise in the same photoionized gas.

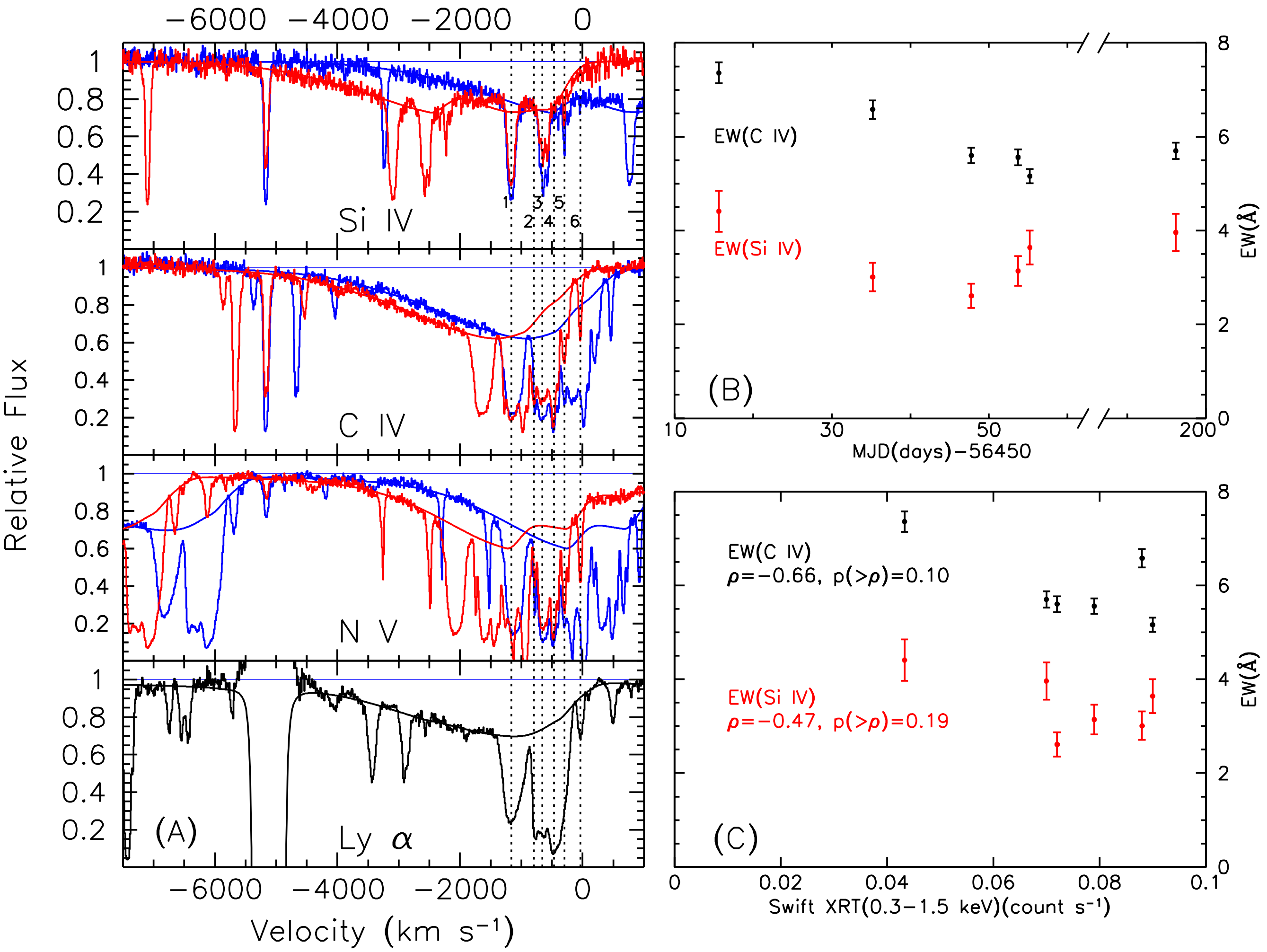

**Fig. 2.** UV absorption lines in the COS spectrum of NGC 5548. (A) The combined normalized spectra from summer 2013 have been binned for clarity. For the Si IV, C IV, and N V doublets, the red and blue profiles are registered relative to the respective red and blue wavelengths of the doublets. The smooth solid lines are fits to the broad absorption line profile with 1:1 ratio for the doublets. The dotted vertical lines show the locations of the narrow velocity components defined in (*20*), with an additional component 6 near 0 velocity.
(B) Time variability of the equivalent width (EW) of the C IV and Si IV broad absorption troughs (including the 2014 measurement). Error bars are ±1 SD.
(C) Equivalent widths of the C IV and Si IV broad absorption troughs, showing an anti-correlation with the Swift soft X-ray flux (0.3−1.5 keV). Error bars are ±1 SD.

As indicated above, the UV manifestation of the persistent classical WA has a significantly lower degree of ionization due to this obscuration. Lower ionization will also cause a significant increase in the X-ray opacity of the WA that needs to be taken into account when modeling the obscuring medium in the X-ray band. Therefore we developed a proper model for the 2013 WA (*5*) based on the physical characteristics of the WA and SED as measured in 2002 (*6*), when there was no obscurer present.

The best-fit model with a single obscuring component produced a very hard continuum spectrum with a photon index of 1.48, inconsistent with the high-energy NuSTAR data that are not affected by obscuration. We required two obscuring components to find a solution consistent with the hard X-ray spectrum (Fig. S3), which has a photon index of 1.57.

The first obscuring component covers 86% of the X-ray source and has a hydrogen column density of $1.2\text{x}10^{26}$ $m^{-2}$ and log $\xi = -1.2$ (in units of $10^{-9}$ Wm). This component, derived from X-ray analysis only, reproduces all the broad absorption troughs seen in the UV. The similar depths of the red and blue transitions for Si IV, C IV, and N V in Fig. 2A indicate that the troughs are saturated. They only partially cover the broad lines and continuum, such that the line profiles indicate the velocity-dependent covering factor rather than the column density profile.

The troughs in C IV and Lyα are deep enough to fully cover the continuum emission, but it is not possible to unambiguously determine the separate broad-line and continuum covering fractions. If the continuum is fully covered, then the covering factor of the C IV BLR is 20%. For N V and Si IV the maximum continuum covering fractions are 95% and 40%, respectively. For continuum covering fractions of 30% in each line, the BLR covering fractions of C IV, N V and Si IV are 30%, 40%, and 20%, respectively. The signal in the RGS soft X-ray band is too low and dominated by features from the narrow emission lines and warm absorber to allow useful constraints on outflow velocity or line width of the obscurer. Therefore the COS UV spectra are essential for understanding the dynamics of the outflow.

The second obscuring component covers 30% of the X-ray source with $N_H = 10^{27}$ $m^{-2}$ and is almost neutral. Taking the same ratio between the X-ray and the UV continuum covering factors as for component 1 (86% for X-ray, 30% for UV), the UV covering factor of component 2 is small (less than 10%).

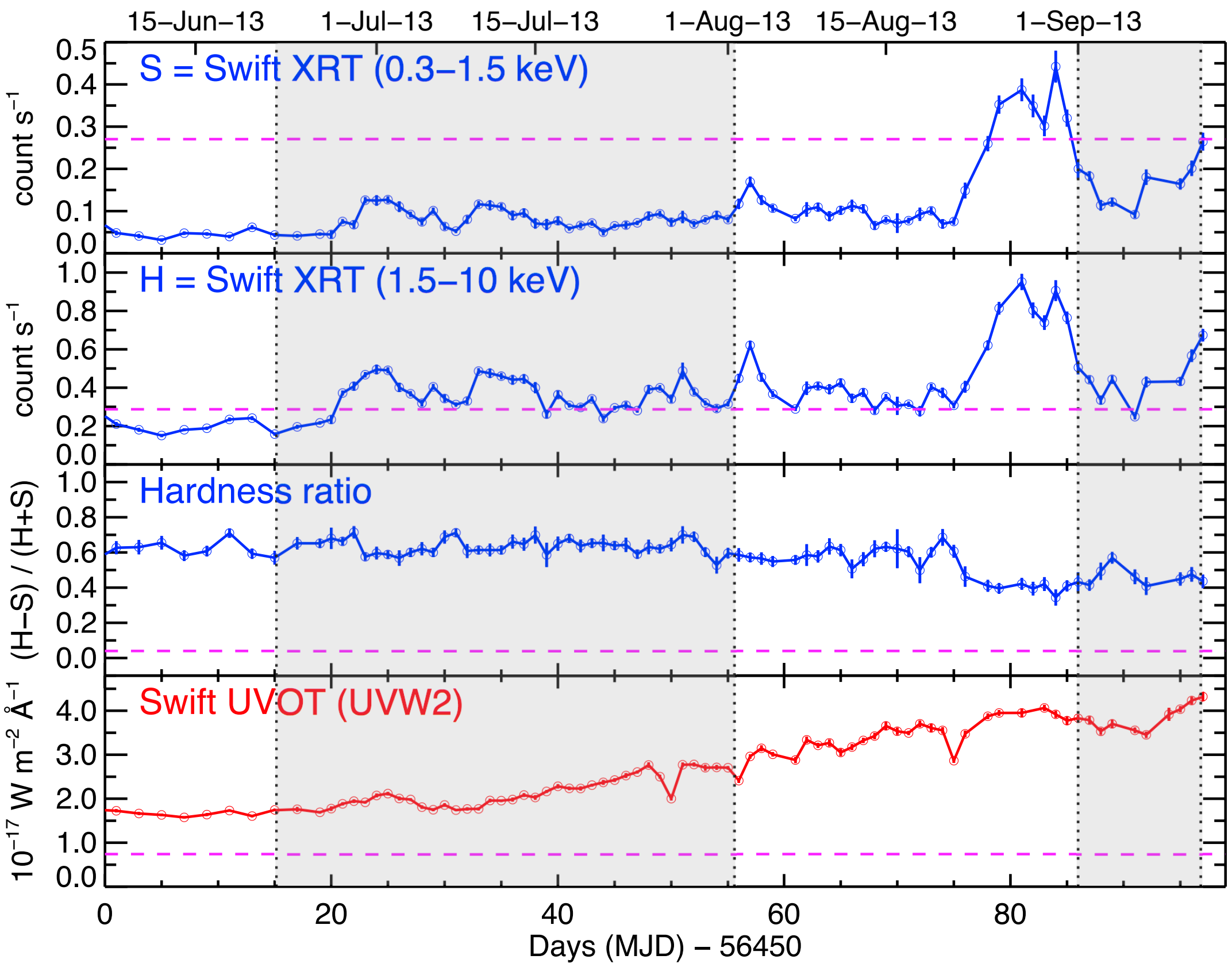

**Fig. 3**. X-ray and UV light curves of NGC 5548 obtained with Swift. Horizontal dashed lines show the average flux or hardness measured with Swift during 2005 and 2007 in an unobscured state. The first and second shaded areas indicate the epochs of the first 12 XMM-Newton and the three Chandra observations, respectively. The outburst around day 80 was also obscured, as evidenced by the hardness ratio. Note the apparent anti-correlation between hardness and UV flux. Error bars are ±1 SD. The full light curve is shown in Fig. S4.

The obscuration is present during our full campaign – see Fig. 3. Archival Swift data (Fig. S4) show large hardness ratios, a strong signature of photoelectric absorption, indicating the obscuration started between August 2007 and February 2012, when Swift was not monitoring NGC 5548; a HST/COS spectrum taken in June 2011 already indicates an onset of the broad absorption lines. Thus the obscuration has lasted for at least 2.5 years, and perhaps as long as six years.

Interestingly, in early September 2013, NGC 5548 brightened for about two weeks (Fig. 3). The Swift X-ray hardness ratio indicated that the source was still obscured, so the continuum itself must have increased. The last Chandra LETGS spectrum of 11 September 2013 showed the effect of the outburst as an increase in the ionization degree of the WA and re-appearance of several X-ray lines from the WA, but no noticeable effect on the obscurer.

What is the obscurer and where is it located (Fig. 4)? The UV troughs are deeper than the continuum, and the obscurer partially covers the broad UV emission lines, implying a distance of more than 2−7 light days ($10^{14}$ m) from the core (*7*). However, it strongly affects the WA by reducing the level of ionization, implying the obscurer is within ~$10^{17}$ m from the core (*20*). Given the UV broad line covering factor of obscurer component 1 of 20–40%, it is likely that the obscurer is close to the broad UV line-emitting region. Based on the UV/X-ray ratio of continuum covering factors determined before for the obscuring component 1, it follows that the UV source is only two times larger than the X-ray source. Another indication for a distance just outside of the broad line region (BLR) is the high velocity of the gas in the line of sight, peaking at −1000 km/s and extending out to −5000 km/s. We see variations in the obscuration between subsequent XMM-Newton observations, spaced by two days. Assuming that the X-ray source has a typical size of $10R_g$ ($R_g = GM/c^2$) and a black hole mass $M = 3.9\text{x}10^7$ solar masses (*8*), the velocity needed to cross this distance in two days is 3000 km/s, of the same order of magnitude as the velocities observed along the line of sight. On the other hand, this obscuration lasts for several years, indicating ongoing replenishment of the obscuring material, possibly from the accretion disk or from the BLR.

Given the location, the outflow velocity, and the changes in the X-ray covering fraction of the obscurer, we may attribute the obscuring material to a wind from the accretion disk reaching beyond the BLR. The line of sight is inclined by ~$30^o$ to the rotation axis of the disk (*8*), and these velocity components therefore indicate a predominantly poloidal outflow. This is more readily explained by magnetically confined acceleration than by radiative acceleration, which would lead to a more radial or equatorial outflow.

Evidence is accumulating for strong variations in the X-ray absorption properties along the line of sight for both type-1 and type-2 AGNs. X-ray absorbers show changes from Compton-thin to Compton-thick columns, changes in covering fraction, eclipse/occultation events, variations in the ionization state, and velocity of some of their major components (*9−17*). Observations so far suggest that these examples of circumnuclear absorbing gas are inhomogeneous and clumpy, preferentially distributed along the equatorial plane, and possibly due to either a clumpy torus or clouds of the BLR passing along the line of sight. On the other hand, the distant WA seems to persist over much longer time scales.

Compared with previous studies, our observations of this type-1 AGN stand out because i) this persistent X-ray/UV obscuring event implies a long, inhomogeneous stream of matter, possibly the onset of an accretion disc wind, and ii) this phenomenon was detected simultaneously in X-rays and UV.

Outflows powerful enough to influence their environments are present in luminous quasars like BAL QSOs. Theoretical models of radiatively driven accretion-disk winds in BAL QSOs require that the accelerated UV-absorbing gas is shielded from much of the UV/X-ray ionizing radiation (*18*). They predict heavy X-ray obscuration and broad UV absorption lines arising from gas near the accretion disk. Observational evidence is ambiguous, as (*19*) see evidence for both X-ray absorption and intrinsically weak X-ray emission in BAL QSOs.

The proximity and brightness of nearby Seyfert galaxies make them ideal laboratories for studying the mechanisms that drive the powerful winds seen in their more luminous cousins.

Our observations of NGC 5548 show that the X-ray obscuration and the fast broad UV outflow both arise in the region expected for an accretion disk wind, and we established that this shielding gas is near the broad-line region.

Although the outflow in NGC 5548 is not strong enough to influence its host galaxy, it gives us unique insight into how the same mechanism may be at work in much more powerful quasars.

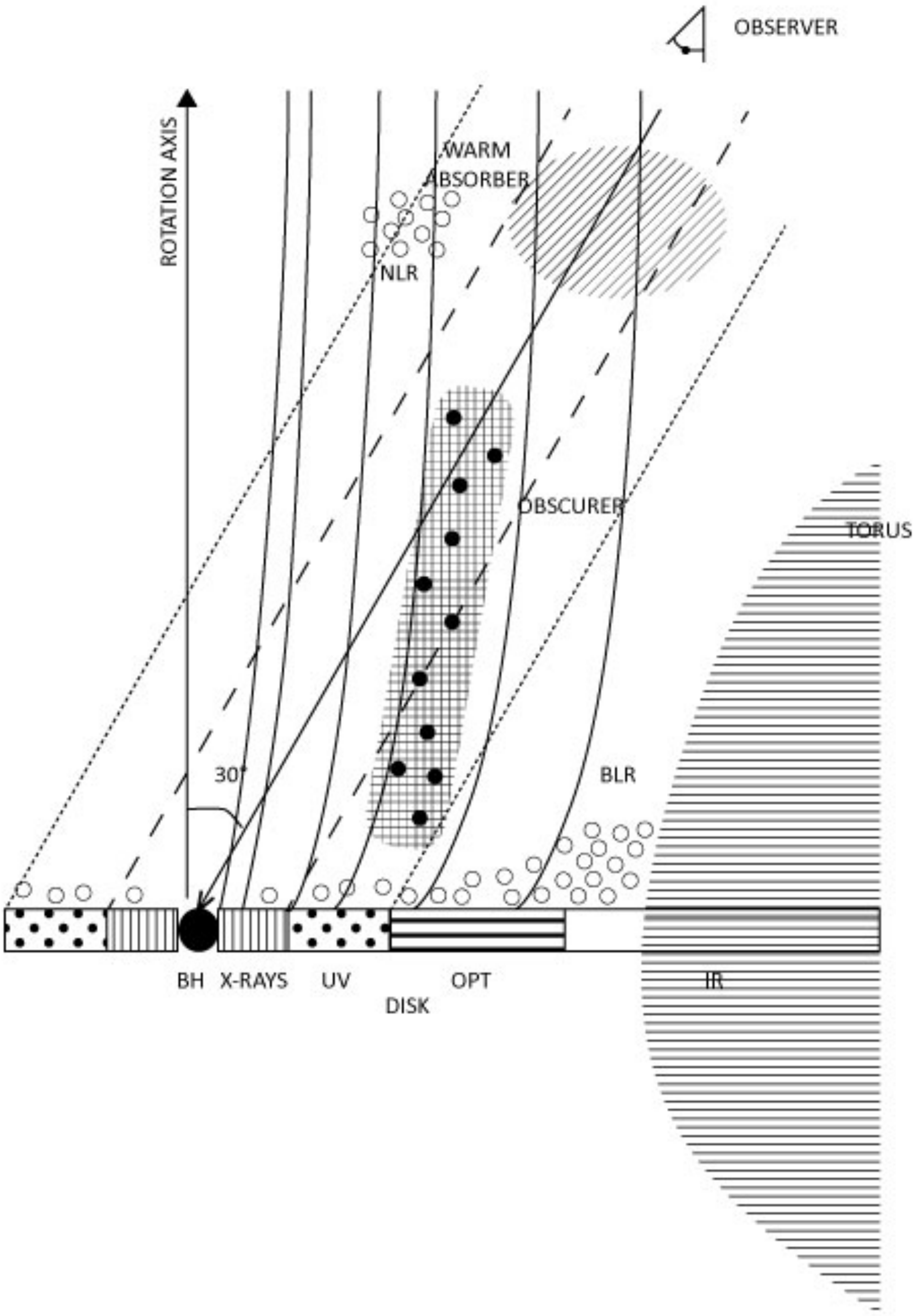

**Fig. 4**. Cartoon of the central region of NGC 5548 (not to scale). The disk around the black hole (BH) emits X-ray, UV, optical and IR continuum and is surrounded by a dusty torus. The curved lines indicate the outflow of gas along the magnetic field lines of an accretion disk wind. The obscurer consists of a mixture of ionized gas with embedded colder, denser parts and is close to the inner UV broad emission line region (BLR). The narrow line region (NLR) and the persistent warm absorber (WA) are farther out.

**References and Notes:**

1. C.S. Reynolds, *MNRAS* **286**, 513-537 (1997).
2. D.M. Crenshaw *et al.*, *ApJ* **516**, 750-768 (1999).
3. J.S. Kaastra *et al.*, *A&A* **534**, A36 (2011).
4. J.S. Kaastra *et al.*, *A&A* **539**, A117 (2012).
5. See supplementary materials on Science Online.
6. K.C. Steenbrugge *et al.*, *A&A* **434**, 569-584 (2005).
7. Korista, K. T., *et al.*, *ApJ Suppl.* **97**, 285 (1995).
8. A. Pancoast *et al.,* arXiv:1311.6475 (2013).
9. G. Risaliti, M. Elvis, F. Nicastro, *ApJ* **571**, 234-246 (2002).
10. G. Lamer, P. Uttley, I.M. McHardy, *MNRAS* **342**, L41-L45 (2003).
11. D.M. Crenshaw, S.B. Kraemer, *ApJ* **659**, 250-256 (2007).
12. T.J. Turner, J.N. Reeves, S.B. Kraemer, L. Miller, *A&A* **483**, 161-169 (2008).
13. S. Bianchi *et al.*, *ApJ* **695**, 781-787 (2009).
14. G. Risaliti *et al., ApJ* **696**, 160-171 (2009).
15. A.L. Longinotti *et al.*, *ApJ* **766**, 104 (2013).
16. G. Miniutti *et al.*, *MNRAS* **437**, 1776-1790 (2014).
17. A.G. Markowitz, M. Krumpe, R. Nikutta, *MNRAS*, **439**, 1403-1458 (2014).
18. D. Proga, T.R. Kallman, *ApJ* **616**, 688-695 (2004).
19. S. Gallagher *et al., ApJ* **644**, 709-724 (2006).
20. D.M. Crenshaw *et al.*, *ApJ* **698**, 281-292 (2009).

**Acknowledgments:**

The data used in this research are stored in the public archives of the international space observatories involved: XMM-Newton, HST, Swift, NuSTAR, INTEGRAL and Chandra. We acknowledge support by ISSI in Bern; the Netherlands Organization for Scientific Research; NASA HST program 13184 from STScI, which is operated by AURA, Inc., under NASA contract NAS5-26555; the UK STFC; the French CNES, CNRS/PICS and CNRS/PNHE; VRIDT (Chile); Israel's MoST, ISF (1163/10), and I-CORE program (1937/12); the Swiss SNSF; the Italian ASI grant INAF I/037/12/0-011/13 and INAF/PICS; US NSF grant AST-1008882; EU Marie Curie contract FP-PEOPLE-2012-IEF-331095. We thank Silvana Paniagua for helping with Fig. 4.

**Supplementary Materials:**

Materials and Methods

Figures S1-S4

Tables S1-S3

Movie S1

References (*21-38*)

**Supplementary Materials:**

**Materials and Methods:**

**1. Observational campaign and data analysis**

We have carried out a large multi-wavelength campaign on NGC 5548 from May to September 2013, and again from December 2013 to February 2014. This unprecedented campaign has utilized five X-ray observatories (XMM-Newton, Swift, NuSTAR, INTEGRAL, and Chandra), the HST and two ground-based optical telescopes (the Wise Observatory and the Cerro Armazones Observatory). At the core of our campaign in summer 2013 (22 June to 1 August), there were 12 XMM-Newton observations, from which five were taken simultaneously with HST COS, four with INTEGRAL and two with NuSTAR observations. The summer observations were followed by three Chandra LETGS observations in the first half of September 2013, with one of them taken simultaneously with NuSTAR. A few months later, two more XMM-Newton observations were taken in December 2013 and February 2014, the former simultaneous with HST COS and NuSTAR observations. There were also two more INTEGRAL observations made in January 2014. Throughout our campaign, Swift monitored NGC 5548 on a daily basis.

In all the XMM-Newton observations (~50−55 ks each), the EPIC instruments (*21, 22*) were operating in the Small-Window mode with the thin-filter applied. The data were processed using the XMM-Newton Science Analysis System (SAS v13). The RGS instruments (*23*) were operated in the Spectro+Q mode. Using accurate relative calibration for the effective area of the RGS and an accurate absolute wavelength calibration (*24*) a fluxed RGS spectrum for each observation was created. Those spectra were then stacked taking into account the effects of the XMM-Newton multi-pointing mode. The RGS spectrum used in this work was produced by stacking the first-order RGS-1 and RGS-2 spectra from the 12 XMM-Newton observations of summer 2013. In each XMM-Newton observation, the OM instrument (*25*) took images in the V, B, U, UVW1, UVM2, and UVW2 broadband filters and also obtained spectra with the Visible and UV Grisms.

In the Swift (*26*) observations (~1−2 ks each), the XRT (*27*) was operating in the Photon Counting mode and the UVOT data (*28*) were taken in the V, B, U, UVW1, UVM2, and UVW2 filters during the summer campaign. Together with other Swift programs, NGC 5548 has been monitored on average every two days in 2013, with UVOT observations taken mostly with the UVW2 filter and the UV Grism. The six HST COS visits (~3700 s each), observing through the Primary Science Aperture, were taken with gratings G130M and G160M, covering the far-UV spectral range from 1132 Å to 1801 Å. The three Chandra observations (30 ks, 67 ks and 122 ks) were taken with LETGS using the HRC-S camera.

The NuSTAR (*29*) observations were reduced with the standard pipeline (NUPIPELINE) in the NuSTAR Data Analysis Software (NUSTARDAS, v1.3.0; part of the HEASOFT distribution as of version 6.14), and instrumental calibration files from NuSTAR caldb v20131223 were used throughout. The raw event files were cleaned with the standard depth correction, which significantly reduces the internal background at high energies. Data taken during passages of the satellite through the South Atlantic Anomaly were removed. Source products were obtained from circular regions (radius ~110"), and background was estimated from a blank area on the same detector. Spectra and lightcurves were extracted from the cleaned event files using NUPRODUCTS for each of the two hard X-ray telescopes (modules A and B) aboard NuSTAR. Finally, the spectra were grouped such that each spectral bin contains at least 50 counts.

The INTEGRAL (*30*) observations were partially simultaneous with the XMM-Newton observations, for a total ON_TIME exposure of about 320 ks. INTEGRAL data were reduced with the Off-line Scientific Analysis software OSA 10.0. IBIS/ISGRI spectra were extracted using the standard spectral extraction toolibis_science_analysis and combined together using the OSA spe_pick tool to generate the total ISGRI spectrum, as well as the corresponding response and ancillary files.

Each HST observation consisted of a two-orbit visit using COS (*31*) in TIME-TAG mode. Following a spectroscopic target acquisition, we obtained four exposures (390–440 s each) with grating G130M at central wavelength settings of 1291 and 1327, using FPPOS=3 and 4 for each. These exposures covered the wavelength range of 1132–1467 Å. We then obtained four exposures (530–550 s each) with grating G160M at central wavelength settings of 1600, 1611, and 1623 at FPPOS=1, and 1623 at FPPOS=4, covering 1416–1801 Å. The multiple central wavelength and FPPOS settings enable us to combine the spectra to bridge the gap between the two COS detector segments, and the multiple focal plane positions also move dead spots and flat-field errors around in wavelength space so that all regions of the spectrum are sampled at least once by clean areas on the detector. In combining the G130M spectra, times corresponding to orbital daylight with wavelengths falling between 1300 and 1310 Å were filtered out of the combination in order to eliminate airglow due to O I emission from Earth's upper atmosphere.

For better wavelength calibration, we cross-correlate the 1998 STIS spectrum with sections of each COS observation containing strong Galactic absorption features to determine any adjustments to the wavelength scale for each COS observation. For each spectrum a simple linear adjustment (amounting to < 20 km/s) sufficed to give a wavelength scale accurate to < 5 km/s (rms). However, there are regions, typically within 100 Å of the edges of each spectrum, where wavelength offsets of up to 20 km/s are observed. We also apply better flat-field corrections and refined flux calibrations that take into account up-to-date adjustments to the time-dependent sensitivity of COS. At wavelengths >1200 Å, we estimate the flux accuracy to be better than 2%; at shorter wavelengths, the absolute calibration error may be up to 5%. The applied pixel-to-pixel flat fields support S/N ratios of 30:1 per pixel or better, Poisson statistics permitting.

For the average spectrum used to produce the normalized absorption line spectra in Fig. 2, each individual observation from the summer of 2013 was combined using a weighting of exposure time times the flux at 1360 Å (G130M) and 1485 Å (G160M). This approximates optimally weighting each spectrum by its total counts, as appropriate for Poisson-distributed photon-arrival rates. This combined spectrum was then binned on four-pixel intervals (approximately one-half a COS resolution element) to enhance the signal-to-noise ratio without sacrificing spectral resolution. Likewise, the individual spectra analyzed in Fig. 2A were binned at intervals of four pixels.

The Wise Observatory monitoring of NGC 5548 was done with the Centurion 46-cm telescope, taken in the standard B, V, R, and I filters. They cover the time from beginning of June to end of September 2013. The Cerro Armazones Observatory monitored NGC 5548 with the 25-cm Berlin Exoplanet Search Telescope II between 2 May and 25 July, taking images in the B, V and R bands.

## 2. Analysis of the UV broad emission and absorption lines

The C IV and Lyα emission lines in our COS spectra of NGC 5548 show very asymmetric profiles. Even after allowing for the blueshifted narrow absorption lines, the blue side of each line profile appears depressed compared to the red side. This depression of the blue side of the line profile could also be an intrinsic asymmetry in the emission-line profile, but three main arguments support broad, blueshifted absorption:

1. The depressions appear in association with all resonance lines and low-excitation permitted transitions in the spectrum. Notably, no depressions are seen on the blue wings of He II λ1640, O III] λ1663, and N III] λ1750, and N IV] λ1486 (see Fig. 1 bottom and Fig. S1).
2. There are also weak troughs on the blue sides of weaker lines such as C II λ1335, Si II λ1302, C III* λ1176, and the P V and Fe III transitions in the 1130–1150 Å observed wavelength range. These troughs mainly affect the continuum, as there is little to no line emission associated with these features that can be modeled with a blue/red emission asymmetry (see Fig. 1 bottom).
3. By modeling the depressions as absorption, we can fit all strong emission lines with a similar profile that is largely symmetric. In addition, this same symmetric emission model can fit the historical spectra with only slight adjustments in parameters. If the depression is modeled as an emission asymmetry, drastically different line profiles are required for Lyα, C IV, and He II, even in the same COS spectrum, and these profiles must change dramatically with respect to the archival spectra (Fig. S2).
4. The changes in the asymmetry of the broad UV emission lines due to absorption show a weak correlation with the opacity of the absorber in the soft X-ray spectrum, indicating a relation between them (Fig. 2C).

To model the UV emission from NGC 5548, for the continuum we used a reddened power law (with extinction fixed at E(B-V) = 0.02 mag) plus weak Fe II emission longward of 1550 Å rest wavelength (*32*), broadened by 4000 km/s (FWHM), absorbed by a Galactic damped Lyα absorber with fixed $N_H = 1.45 \times 10^{24}$ $m^{-2}$ (*33*) and centered at –13 km/s (based on H I emission). For the brightest emission lines (Lyα, N V, Si IV, C IV, and He II) we use five Gaussian components — narrow line emission (NLR; ~300 km/s FWHM), intermediate line emission (ILR; ~860 km/s), modestly broad emission (~2500 km/s), broad emission (~8500 km/s), and very broad emission (~16,000 km/s). For N V, Si IV and C IV, each doublet component is modeled separately for the first four of these Gaussians. The relative intensity ratios are fixed at 1:1, assuming the emission is optically thick. Only a single component is used for the very broad emission. For Lyα, C IV, and He II, the best-fit values for the NLR and ILR components are very similar to those in the low-state STIS spectrum of 2004, e.g., see (*20*). For all other lines, we fix the intensities and widths of the ILR and NLR components at values measured in the 2004 STIS spectrum since these weaker lines cannot be reliably measured independently in our COS spectrum due to their weakness relative to the much brighter continuum and surrounding broader emission.

The broad absorption troughs in the spectrum have a noticeable asymmetry. The troughs have a much greater extent to velocities blueward of their deepest point than they do to the red. To model these troughs, we used asymmetric Gaussians with a negative flux profile in which the half-width at half maximum (HWHM) of the blue side of the trough was larger than the HWHM of the profile of the red side. This asymmetric absorption profile was used for every permitted transition in the spectrum. For the doublets of N V, Si IV, and C IV, which are largely unresolved, we fixed the relative depths of the red and blue transitions at an optically thick ratio of 1:1. The red transition of each doublet was constrained to have the same width and velocity as the blue transition. In addition, since the N V broad absorption profiles were not well constrained, their width and velocity were fixed at the values obtained for C IV.

We then fit this parameterized emission model to the combined NGC 5548 spectrum excluding regions affected by foreground Galactic absorption lines and the narrow intrinsic absorption lines in NGC 5548, but we optimized the parameters describing the broad absorption profiles simultaneously with those of the line and continuum emission. Fig. S1 illustrates our fit to the region surrounding the C IV emission line in NGC 5548. Table S1 gives the best-fit parameters for the broad absorption lines associated with Lyα, N V, Si IV, and C IV.

We assume that the absorbing gas covers all emission components except for the NLR emission, cf. (*20*). To normalize the spectra shown in Fig. 2A, we then divide the observed spectrum by the emission model including all components except those describing the NLR (see also Fig. S1).

### 3. Re-analysis of the un-obscured X-ray spectrum of NGC 5548 in 2002

In order to characterize the WA in its unobscured state we re-analyzed the Chandra spectrum of 2002. The LETGS spectrum and response matrix were exactly the same as used by (*6*), and only data in the 2–60 Å range were used. For the HETGS spectrum we took the data from the *TGCat* archive (*34*), and we used the 1.55−15.5 Å range for the HEG and 2.5−26 Å for the MEG. To get consistent results between the HEG and MEG grating, the MEG fluxes were scaled by a factor of 0.954 relative to the HEG fluxes.

We used the SPEX package (*35, 36*) for spectral analysis. As in (*6*), we used a power-law plus modified blackbody spectrum for the continuum. We added four broad emission lines representing the 1s−2p and 1s−3p transitions of O VII and the 1s−2p transition of O VIII and C VI, with their parameter values taken from (*6*).

The continuum and broad lines are absorbed by six WA ionization components. Each WA component is represented by the *xabs* model of SPEX, with free parameters the hydrogen column density $N_{\mathrm{H}}$, the ionization parameter $\xi$, the outflow velocity $v$, and the Gaussian turbulent velocity $\sigma_v$. The covering factor of all WA components is unity and the abundances are proto-solar (*37*). We used the same ionizing spectral energy distribution as used by (*6*), Fig. 1. However in contrast to that work, we used the latest version of Cloudy, Version 13.01 (*38*) to calculate the ion concentrations for each value of $\xi$.

In addition to the absorbed continuum and broad lines, we also included a set of narrow emission lines in our model as well as radiative recombination continua (RRCs). The lines were modeled using Gaussians, with the same velocity dispersion $\sigma_v$ for each component. Also the RRCs were convolved with this velocity dispersion, but we assumed that the WA did not absorb these narrow components. The parameters from these components were taken from a preliminary fit to the obscured RGS spectrum of 2013, where the lines and RRCs are clearly visible due to the strong obscuration of the continuum source. In the 2002 spectrum, most lines are difficult to detect due to the strong continuum; therefore we kept their normalizations fixed to the 2013 values, except for the strong forbidden [O VII] line.

All emission components were corrected for a cosmological redshift $z = 0.017175$ and Galactic absorption from cold material using the *hot* model of SPEX, with a nominal temperature of 0.5 eV, proto-solar abundances, and a total hydrogen column density $N_H = 1.45 \times 10^{24}$ m$^{-2}$ (*33*).

We use C-statistics and report errors at the 1σ (68%) confidence level. The HETGS and LETGS spectra were jointly fitted but allowed different values of the power-law and modified blackbody component for each spectrum, because NGC 5548 showed significant variability during these non-simultaneous observations. After optimizing this fit, we allowed the ionization parameters of the six WA components to be different for each spectrum, in order to account for possible changes between the two observations, but all other WA parameters were kept fixed.

The best-fit parameters of this model are shown in Table S1.

**4. Analysis of the obscured X-ray spectrum of NGC 5548 in 2013**

We analyzed the stacked spectrum of the 12 XMM-Newton observations taken during summer 2013. We used the RGS data between 5.68 and 38.23 Å. We only used the first spectral order and stacked the data of both RGS detectors. Further we used the data from the pn detector in the range between 1.24 and 12 keV. Starting in 2013, a gain problem with the pn data appeared, causing a shift of about 50 eV near the Fe-K complex. This small offset yields poor fits near the energy of the gold M-edge of the telescope mirror, and for that reason the interval between 2.0 and 2.4 keV was omitted from the fit.

The dynamics of the WA has hardly changed over 20 years, so variations in the WA are due to ionization changes rather than changes in its hydrogen column density.

This allows us to derive the properties of the obscuring medium by iteration. Because the ionizing continuum at high energies in 2013 is very similar to that in 2002, we initially adopt as our model the 2002 continuum plus WA. The difference with respect to the observed average spectrum is attributed to absorption by the new obscurer. The obscured continuum is then used as the ionizing source for the WA. The relative ion concentrations and opacities of the WA components are then re-evaluated using the new obscured spectral energy distribution and ionizing luminosity. This process is iterated a few times until it converges. Next, the continuum parameters are left free and a few more iterations are made.

The spectroscopic tools and models are exactly the same as for the unobscured 2002 spectrum, with the following exceptions.

We replace the power-law model with the *refl* model of SPEX, which includes, in addition to the power-law, reflection by cold material and an exponential cut-off $E_c$ at high energies. The cut-off energy was determined from a preliminary fit to the combined pn, NuSTAR and INTEGRAL spectra, but kept frozen in the present analysis of the RGS and pn spectra. Also, we applied an artificial blueshift of −0.0126 ± 0.004 to the *refl* component to account for the gain error in the pn, that affects the location of the narrow Fe-K emission line. We kept the reflection component cold and did not apply broadening by a disk. Thus, the reflection component represents distant material. We kept the parameters of the four broad emission lines of oxygen and carbon the same as for the 2002 spectrum.

In addition to the WA, we applied also two obscurer components to the continuum and broad lines, also modeled using the *xabs* model of SPEX that is used for the WA. However, for the obscuring components we used the actual ionizing continuum as represented by the reflection component, the modified blackbody and the broad emission lines, smoothly joined to the same ionizing SED as used for the 2002 spectrum in the EUV band. Because we find that the obscurer is close to the central source, we have cut-off the ionizing continuum in the optical band (i.e., no significant Compton scattering from infrared torus photons).

The ionization parameters of the lower-ionization WA were calculated from the 2002 values by simply scaling with the relevant ionizing luminosities. The values for log $\xi$ in 2013 were lower by 0.45 dex, which is less than the soft X-ray and EUV flux depression by a factor of ten; this is because the ionizing luminosity is determined over the 13.6 eV to 13.6 keV band (1−1000 Ryd), and the hard X-ray flux is not suppressed.

The normalization of the pn with respect to RGS was determined from the overlapping energy bands and was found to be 1.027. The best fit is derived iteratively as described in the main body of the text, and the best-fit parameters of our model are listed in Table S2. The C-stat values show that the fit is not statistically perfect, but it gives an excellent overall description of the observed spectrum. Remaining residuals are attributed to minor imperfections in the instrument calibration and the spectral models used.

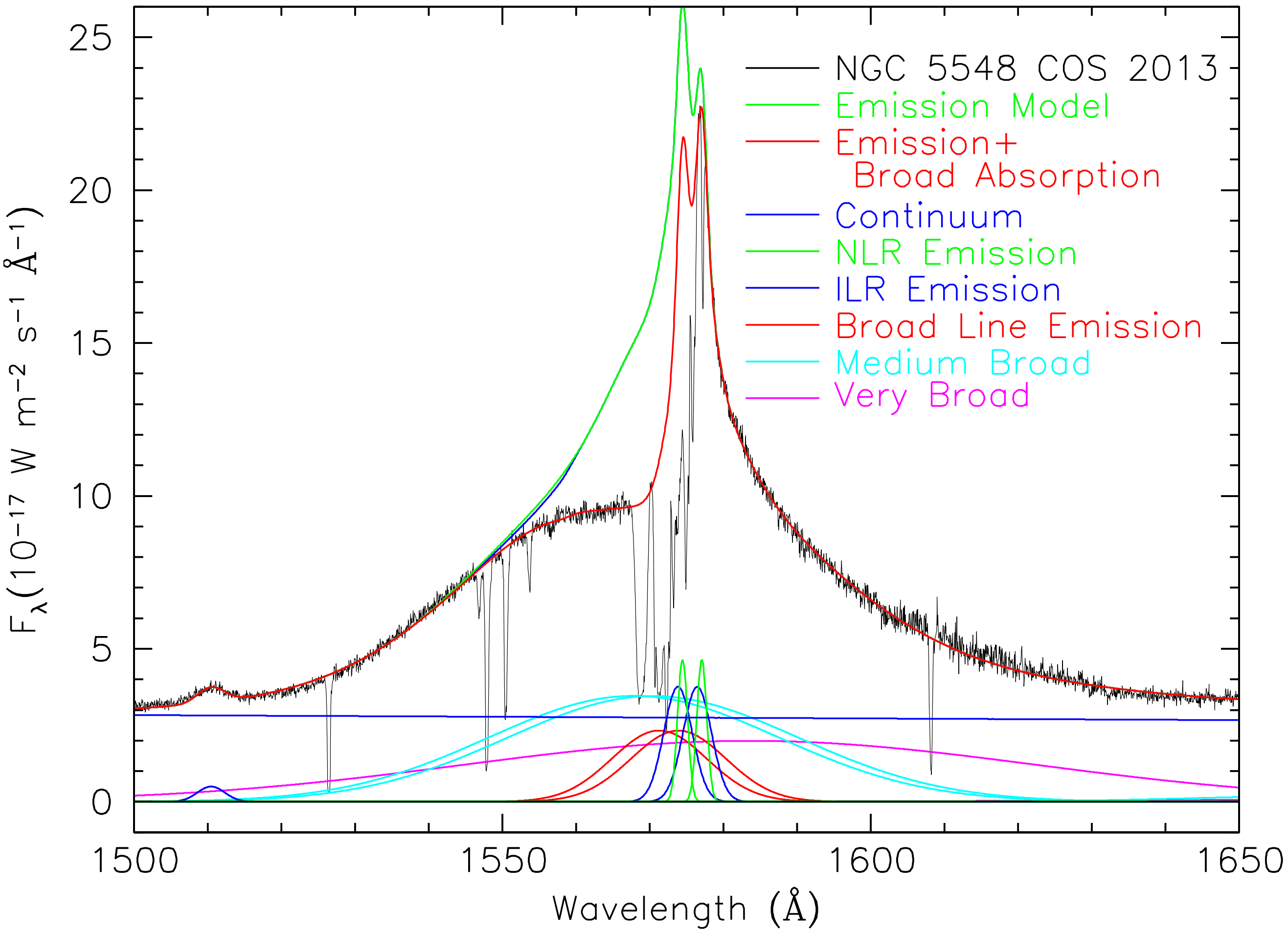

**Fig. S1**. Best-fit emission model to the region surrounding the C IV emission line in the average spectrum of NGC 5548 from summer 2013. The data (black) have been binned by 4 pixels, as in Fig. 2A. The best-fit model (red) includes broad absorption (but excludes narrow absorption and Galactic features). The dark blue solid line is the continuum. The solid green line shows the emission model with the broad absorption removed. The emission components comprising the emission lines are plotted separately below the data.

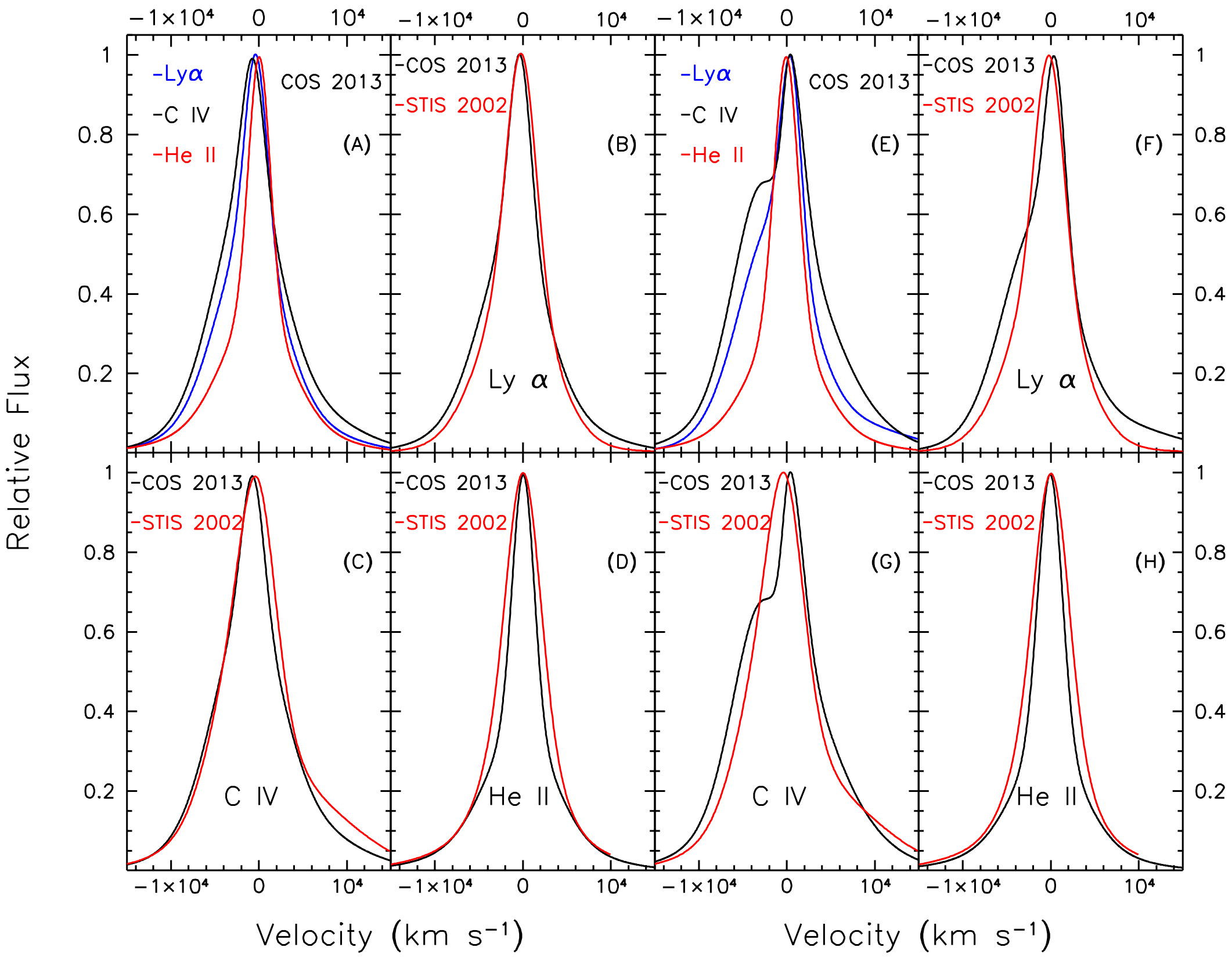

**Fig. S2**. Model UV emission line profiles in NGC 5548. Panels A-D: model line profiles that include broad absorption covering the blue wings of C IV and Lyα. Panel A compares Lyα, C IV and He II line profiles among themselves for the COS 2013 composite spectrum. The panels B-D compare them individually to their counterparts in the STIS 2002 spectrum. These summed profiles omit the NLR and ILR portions of the profiles. Intensities have been rescaled to 1.0 at line center. Panels E-H: Model line profiles that use an asymmetric emission profile instead of broad absorption covering the blue wings of C IV and Lyα.

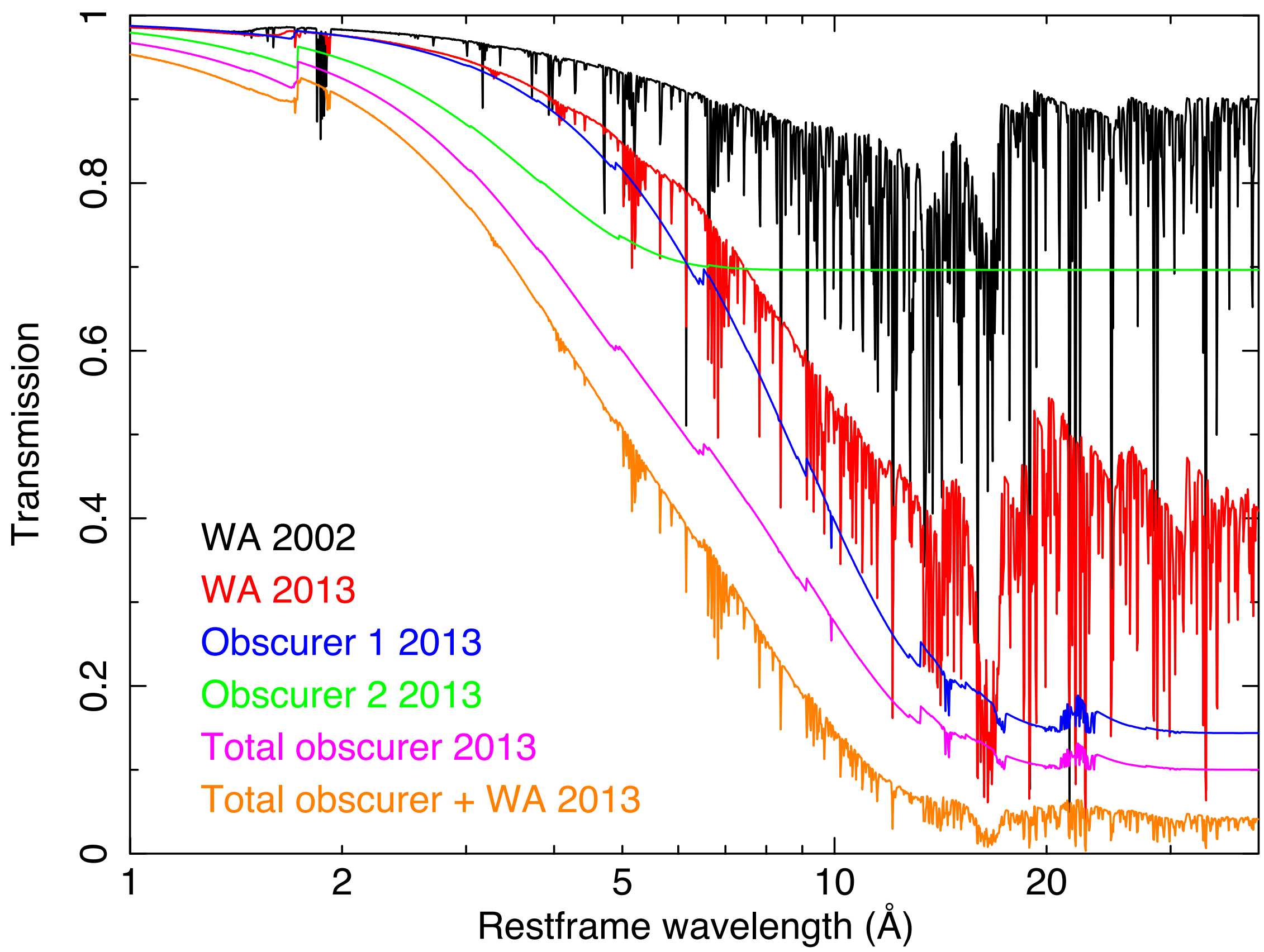

**Fig. S3**. Transmission of the warm absorber (WA) and the obscurer components in 2013. For comparison, the WA transmission for the unobscured state in 2002 is also shown.

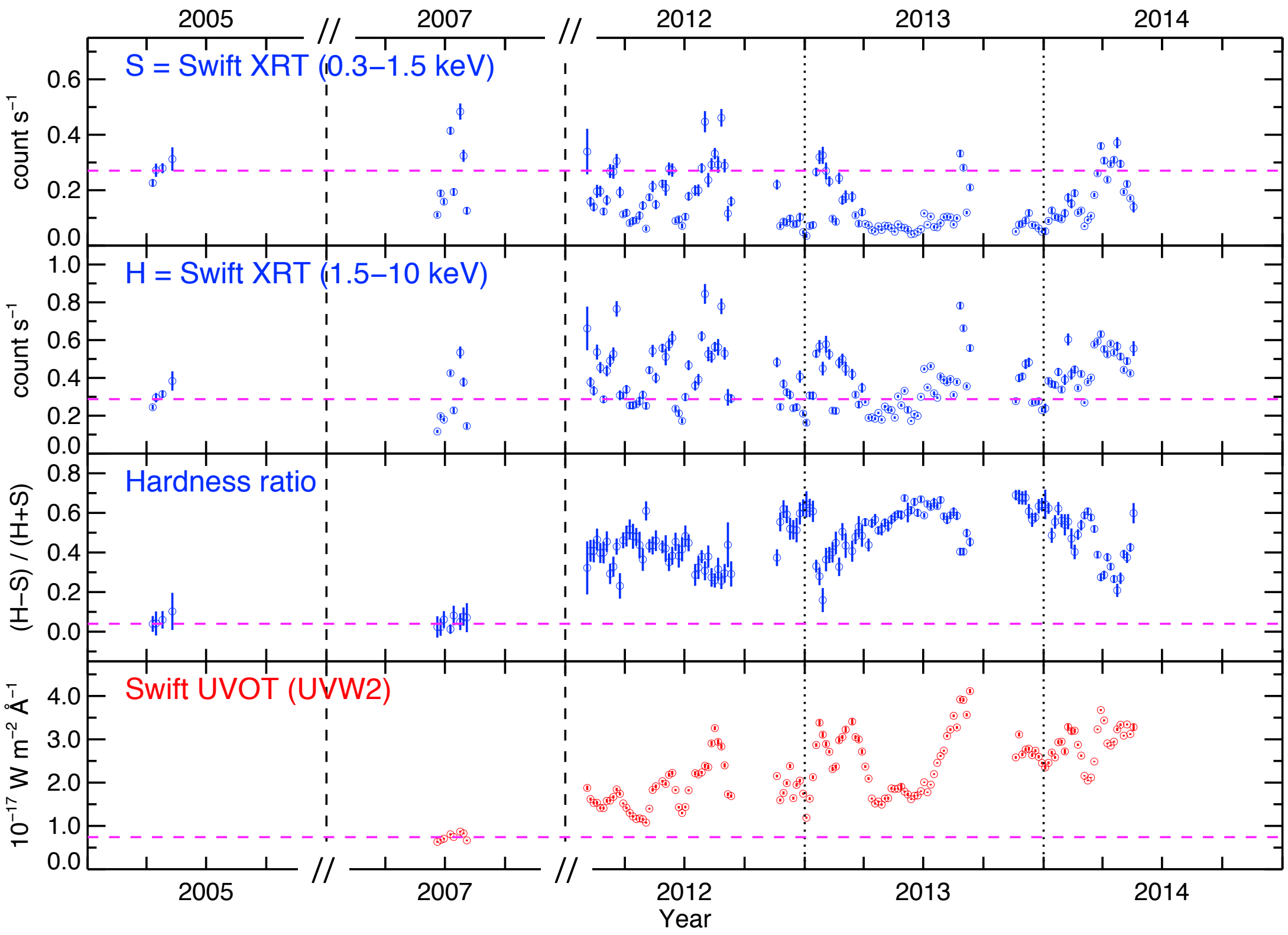

**Fig. S4**. X-ray and UV light curves of NGC 5548 obtained with Swift between 2005 and 2014. Vertical lines separate the five years shown here. Horizontal dashed lines show the average flux or hardness ratio measured with Swift during 2005 and 2007 in an unobscured and optically low state. The data have been averaged over five days. Error bars are ±1 SD. Fig. 3 in the main text shows the light curve during summer 2013 with one-day time-resolution. The hardness ratio, an indicator of absorption, is defined as *(H−S)/(H+S)* with *H* and *S* the count rates in the 1.5−10 and 0.3−1.5 keV bands, respectively. The enhanced hardness ratio during 2012–2014, despite "normal" hard (1.5–10 keV) fluxes, shows that the obscuration is long-lasting.

**Table S1** Best-fit parameters for the broad absorption lines in the COS spectrum of NGC 5548. Velocities are relative to the systemic redshift. Uncertainties are ±1 SD. Numbers without uncertainties are kept frozen.

| Line | $\lambda_o$ (Å) | Equivalent Width (Å) | Velocity (km/s) | $HWHM_{blue}$ (km/s) | $HWHM_{red}$ (km/s) |
|---|---|---|---|---|---|
| Lyα | 1215.67 | 3.26±0.06 | –509±18 | 1923±112 | 337±100 |
| N V | 1238.82 | 1.43±0.03 | –500 | 1883 | 333 |
| N V | 1242.80 | 1.43 | –500 | 1883 | 333 |
| Si IV | 1393.76 | 1.60±0.02 | –535±44 | 1744±91 | 299±32 |
| Si IV | 1402.77 | 1.60 | –535 | 1744 | 299 |
| C IV | 1548.20 | 2.90±0.10 | –500±10 | 1883±100 | 333±27 |
| C IV | 1550.77 | 2.90 | –500 | 1883 | 333 |

**Table S2** Best-fit parameters for the Chandra LETGS and HETGS spectrum from 2002. Numbers without uncertainties are kept frozen.

| Comp. | Parameter | LETGS | | HETGS | | | |
|---|---|---|---|---|---|---|---|
| Power-law | 0.2−10 keV luminosity (W) | $(4.21\pm0.05)\times10^{36}$ | | $(3.66\pm0.05)\times10^{36}$ | | | |
| | Photon index | 1.704±0.013 | | 1.653±0.010 | | | |
| Modified blackbody | 0.2−10 keV luminosity (W) | $(3.16\pm0.22)\times10^{35}$ | | $(6.1\pm2.6)\times10^{35}$ | | | |
| | k$T$ (keV) | 0.140±0.004 | | 0.115±0.010 | | | |
| Broad emission lines | Line: | C VI 1s−2p | O VII 1s−2p | O VII 1s−3p | O VIII 1s−2p | | |
| | Luminosity ($10^{33}$ W) | 1.9 | 3.4 | 1.4 | 2.8 | | |
| | Rest frame wavelength (Å) | 33.74 | 21.6 | 18.63 | 18.97 | | |
| | FWHM (Å) | 0.90 | 0.58 | 0.50 | 0.51 | | |
| Radiative Recom-bination Continua | See Table S3 (very weak compared to continuum) | | | | | | |
| Narrow emission lines | Line | O VII f | C V f | Other lines see Table S3 (very weak compared to continuum) | | | |
| | Wavelength (Å) | 22.101 | 41.472 | | | | |
| | $L$ ($10^{33}$ W) | 4.7±0.4 | 2.3±1.0 | | | | |
| | $\sigma_v$ (km/s) | 382 | | | | | |
| Warm absorber | Component | A | B | C | D | E | F |
| | $N_H$ ($10^{24}$ m$^{-2}$) | 2.0±0.6 | 7.0±0.9 | 15±3 | 10.7±1.6 | 28±8 | 57±17 |
| | LETGS log ξ ($10^{-9}$ Wm) | 0.78±0.08 | 1.51±0.05 | 2.15±0.03 | 2.36±0.03 | 2.94±0.08 | 3.13±0.05 |
| | HETGS log ξ ($10^{-9}$ Wm) | 0.88±0.16 | 1.67±0.06 | 2.42±0.10 | 2.51±0.05 | 2.89±0.05 | 3.35±0.04 |
| | $v$ (km/s) | −588±34 | −547±31 | −1148±20 | −254±25 | −792±25 | −1221±25 |
| | $\sigma_v$ (km/s) | 210±40 | 61±15 | 19±6 | 68±14 | 24±12 | 34±13 |
| Statistics | C-stat = 2402 (LETGS) and 2890 (HETGS) for expected value of 2105 and 2632 | | | | | | |

**Table S3** Best-fit parameters for the co-added 12 XMM-Newton spectra from Summer 2013. Numbers without uncertainties are kept frozen.

| Comp. | Parameter | | | | | | | | |
|---|---|---|---|---|---|---|---|---|---|
| Power-law | 0.2−10 keV luminosity (W) | $(3.63\pm0.08)\times10^{36}$ | | | | | | | |
| | Photon index | 1.566±0.009 | | | | | | | |
| | Reflection fraction | 0.451±0.012 | | | | | | | |
| | $E_c$ (keV) | 184 | | | | | | | |
| Modified blackbody | 0.2−10 keV luminosity (W) | $(5.5\pm1.1)\times10^{35}$ | | | | | | | |
| | k$T$ (keV) | 0.164±0.016 | | | | | | | |
| Broad em. lines | See Table S2 | | | | | | | | |
| Radiative Recom-bination Continua | Ion | C VI | C VII | N VII | N VIII | O VIII | O IX | Ne X | Ne XI |
| | $Y$ ($10^{65}$ m$^{-3}$) | 90±13 | 61±6 | 11±5 | 0±2 | 29±2 | 7.3±1.0 | 4.3±0.9 | 1.6±0.7 |
| | k$T$ (eV) | 5.6±0.4 | | | | | | | |
| Narrow emission lines | Line | Ne X 1s−2p | Ne IX r | Ne IX i | Ne IX f | O VIII 1s−3p | | | |
| | λ (Å) | 12.134 | 13.447 | 13.553 | 13.699 | 16.006 | | | |
| | $L$ ($10^{33}$ W) | 0.09±0.07 | 0.22±0.06 | 0.14±0.05 | 0.78±0.06 | 0.27±0.04 | | | |
| | Line | O VIII 1s−2p | O VII 1s−3p | O VII r | O VII i | O VII f | | | |
| | λ (Å) | 18.969 | 18.628 | 21.602 | 21.807 | 22.101 | | | |
| | $L$ ($10^{33}$ W) | 1.13±0.05 | 0.24±0.04 | 1.09±0.09 | 1.07±0.10 | 3.74±0.14 | | | |
| | Line | N VII 1s−2p | N VI r | N VI i | N VI f | C VI 1s−2p | | | |
| | λ (Å) | 24.781 | 28.787 | 29.084 | 29.534 | 33.736 | | | |
| | $L$ ($10^{33}$ W) | 0.33±0.05 | 0.26±0.06 | 0.25±0.06 | 0.60±0.07 | 1.48±0.09 | | | |
| | $\sigma_v$ (km/s) | 463±26 | | | | | | | |
| Warm absorber | Component | A | B | C | D | E | F | | |
| | log ξ ($10^{-9}$ Wm) | 0.33 | 1.06 | 1.70 | 1.91 | 2.48 | 2.67 | | |
| | $N_H$, v, $\sigma_v$ see Table S2 | | | | | | | | |
| Obscurer | | Component 1 | Component 2 | | | | | | |
| | $N_H$ ($10^{26}$ m$^{-2}$) | 1.21±0.03 | 9.6±0.5 | | | | | | |
| | log ξ ($10^{-9}$ Wm) | -1.20±0.08 | $< -2.1$ | | | | | | |
| | $f_{cov}$ | 0.86±0.02 | 0.30±0.10 | | | | | | |
| Statistics | C-stat = 1845 (RGS) and 433 (pn) for expected value of 1088 and 269 | | | | | | | | |

**Additional references:**

21. L. Strüder *et al.*, *A&A* **365**, L18-L26 (2001).

22. M.J.L. Turner *et al.*, *A&A* **365**, L27-L35 (2001).

23. J.W. den Herder *et al.*, *A&A*, **365**, L7-L17 (2001).

24. J.S. Kaastra *et al.*, *A&A* **534**, A37 (2011).

25. K.O. Mason *et al.*, *A&A* **365**, L36-L44 (2001).

26. N. Gehrels *et al.*, *ApJ* **611**, 1005-1020 (2004).

27. D.N. Burrows *et al.*, *SSRv* **120**, 165-195 (2005).

28. P.W.A. Roming *et al.*, *SSRv* **120**, 95-142 (2005).

29. F.A. Harrison *et al.*, *ApJ* **770**, 103 (2013).

30. C. Winkler *et al.*, *A&A* **411**, L1-L6 (2003).

31. J.C. Green *et al.*, *ApJ* **744**, 60 (2012).

32. B.J. Wills, H. Netzer, D. Wills, *ApJ* **288,** 94-116 (1985).

33. B.P. Wakker, F.J. Lockman, J.M. Brown, *ApJ* **728**, 159 (2011).

34. TGCat data archive, http://tgcat.mit.edu.

35. J.S. Kaastra, R. Mewe, H. Nieuwenhuijzen, in *11th Colloquium on UV and X-ray Spectroscopy of Astrophysical and Laboratory Plasmas*, K. Yamashita, T. Watanabe, Ed. (Univ. Acad. Press, Tokyo, 1996), pp. 411-414.

36. SPEX website www.sron.nl/spex

37. K. Lodders, H. Palme, *Meteoritics and Planetary Science Supp*, 5154 (2009).

38. G. J. Ferland *et al., Rev. Mexic. de Astron. y Astrof.*, **49**, 1–27 (2013).